\documentclass{ivs2e}       % for LaTeX2e class file
\usepackage{graphicx}
\usepackage{epstopdf}
\usepackage{wrapfig} 
\usepackage{color} 
\usepackage[font={footnotesize,up,singlespacing}]{caption}
\Title{Warkworth 12-m VLBI Station: WARK12M}
\STitle{Warkworth Observatory}
\PublicationType{report}
\PublicationName{IVS 2012 Annual Report}
\IvsComponentName{Warkworth Observatory}
\NumberofAuthors{5}
\AuthorName{1}{Stuart Weston}
\AuthorName{2}{Hiroshi Takiguchi}
\AuthorName{3}{Tim Natusch}
\AuthorName{4}{Lewis Woodburn}
\AuthorName{5}{Sergei Gulyaev}

\ContactAuthorReference{1}
\AuthorEmail{1}{stuart.weston@aut.ac.nz}
\AuthorTelephone{1}{+64 9 921 8689}
\AuthorPersonalWebPage{1}{n.a.}
\AuthorEmail{2}{hiroshi.takiguchi@aut.ac.nz}
\AuthorTelephone{2}{+64 9 921 8689}
\AuthorPersonalWebPage{2}{n.a.}
\AuthorEmail{3}{tim.natusch@aut.ac.nz}
\AuthorTelephone{3}{+64 9 921 8689}
\AuthorPersonalWebPage{3}{n.a.}
\AuthorEmail{4}{lwoodbur@aut.ac.nz}
\AuthorTelephone{4}{+64 9 921 8689}
\AuthorPersonalWebPage{4}{n.a.}
\AuthorEmail{5}{sergei.gulyaev@aut.ac.nz}
\AuthorTelephone{5}{+64 9 921 9541}
\AuthorPersonalWebPage{5}{n.a.}
\AuthorInstitutionReference{1}{1}
\AuthorInstitutionReference{2}{1}
\AuthorInstitutionReference{3}{1}
\AuthorInstitutionReference{4}{1}
\AuthorInstitutionReference{5}{1}

\NumberofInstitutions{1}
\InstitutionName{1}{Institute for Radio Astronomy and Space Research, Auckland University of Technology}
\InstitutionPostAddress{1}{Private Bag 92006, Auckland 1142}
\InstitutionCountry{1}{New Zealand}
\InstitutionFax{1}{n.a.}
\InstitutionWebPage{1}{http://www.irasr.ac.nz}

\begin{document}

%
% The command to make the title page from keywords in the header.
%
\MakeTitle                % not \maketitle

\setlength{\parindent}{0.5cm}
\setlength{\parskip}{0.1cm}
%%%%%%%%%%%%%%%%%%%%%%%%%%%%%%%%%%%%%%%%%%%%%%%%%%%%%%%%%%%%%%%
%
% Put your abstract text here.
%
\begin{abstract}
\noindent
The Warkworth 12-m Radio Telescope is operated by the Institute for Radio Astronomy and Space Research (IRASR) 
at AUT University, Auckland, New Zealand. Here we review the characteristics of the 12-m VLBI station and report on a number of activities and technical developments in 2012.
\end{abstract}

\section{General Information}
% Section 1: General Information. Please provide general information about your component such as its location, its %sponsoring agency and what type of contribution you are making to IVS. If you have a photograph of your antenna, %correlator, analysis center, etc. you should include it.
%\vspace{-1cm}

%\begin{wraptable}{l}{95mm}
% \begin{center}
\begin{table}[h]
   \centering
   \caption{Specifications of the Warkworth 12-m antenna.}
%   \par\medskip\par
   \begin{tabular}{ll}
        \hline
        Antenna type & Dual-shaped Cassegrain \\
        Manufacturer & Cobham/Patriot, USA \\
        Main dish Diam.      & 12.1 m  \\
        Secondary refl. Diam. & 1.8 m \\
        Focal length              & 4.538 m \\
        Surface accuracy          & 0.35 mm \\
        Pointing accuracy         & $18''$  \\
        Mount                     & alt-azimuth \\
        Azimuth axis range        & $90^\circ \pm 270^\circ$ \\
        Elevation axis range      & $4.5^\circ$ to $88^\circ$ \\
        Azimuth axis max speed    & $5^\circ$/s \\
        Elevation axis max speed  & $1^\circ$/s \\
        \hline
   \end{tabular}
   \label{t:wark}
\end{table}
% \end{center}
%\end{wraptable}
%        Frequency range           & 1.4---43~GHz \\
%        Main dish F/D ratio:      & 0.375 \\

The WARK12M VLBI station is located some 60 km north of the city of Auckland, near the township of Warkworth. Specifications of the Warkworth 12-m antenna are provided in Table \ref{t:wark}. The radio telescope is equipped with an S/X dual-band dual-circular polarization feed at the secondary focus and an L-band feed at the prime focus. Backend data digitizing is handled by a digital base band converter (DBBC) developed by the Italian Institute of Radio Astronomy. The station frequency standard is a Symmetricom Active Hydrogen Maser MHM-2010 (75001-114). Mark 5B+ and Mark 5C data recorders are used for data storage and streaming of recorded data off site through the network. The observatory network is directly connected to the national network KAREN (Kiwi Advanced Research and Education Network) via a 1 Gbps fiber link to the site \cite{r:karen}.

\section{Component Description}
%Section 2: Component Description. Please provide a technical/scientific description of your component that is relevant to %your component type: for example, parameters of your network station's antenna, capabilities of your correlation center's %correlator, types of analysis solutions that you do as an analysis center, technology developments in progress at your %technology development center, etc.

\subsection{12m: Progress and Issues}

\begin{wrapfigure}{l}{70mm}
 \begin{center}
  \includegraphics[width=70mm, bb=0 0 2048 1536]{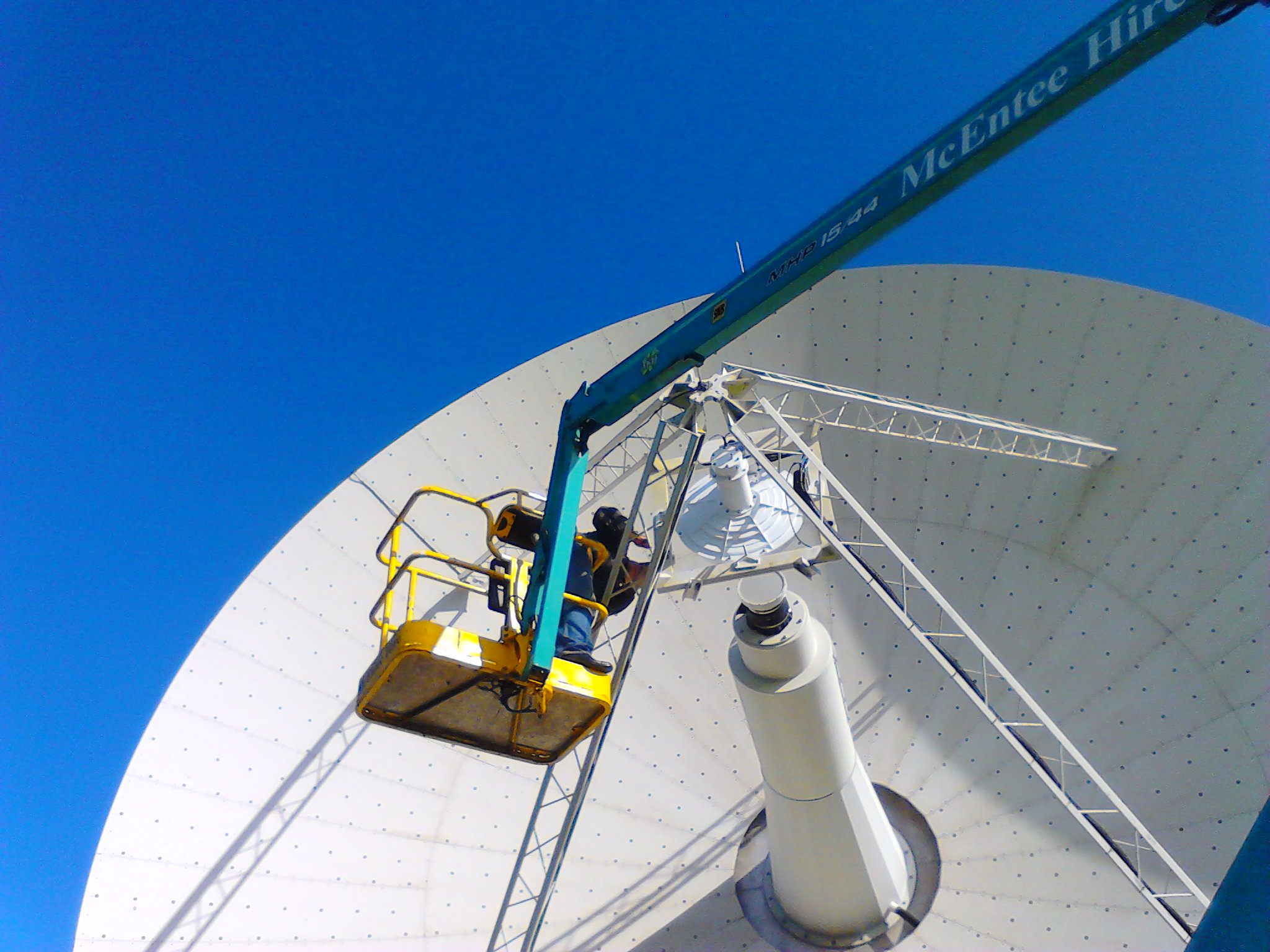}
  \caption{The new L-Band feed installed at prime focus on the 12m. Credit: S. Weston}
% \vspace{-1cm}
  \label{fig:l-receiver}
% \vspace{-2cm}
\vspace{-0.5cm}
 \end{center}
\end{wrapfigure}

The new L-Band feed was designed by Cobham, USA and installed in May 2012 (Figure \ref{fig:l-receiver}). It is located behind the secondary mirror at the prime focus so for L-band observations the secondary mirror has to be removed.

%\begin{figure}[!h]
%\centering
%\begin{tabular}{c}
%\includegraphics[width=7cm, bb=0 0 2048 1536]{S/Install_LBand_12m.jpg}
%\end{tabular}
%\caption{The new L-Band feed installed at prime focus on the 12m. Credit: S. Weston}
%\label{fig:l-receiver}
%\end{figure}

%\vspace{-2.5cm}
\vspace{1cm}

%\vspace{1.8cm}
In late 2012 the Hydrogen maser developed a fault, and it had to be returned to Symmetricom. It is expected to be back and operational in Feb-Mar 2013.
This has severely impacted our stations participation in IVS and LBA observations from late 2012 to date.
Another problem was connected with the jack screw elevation bearing; this was finally repaired in early February 2013. This would appear to have been caused by poor initial assembly with the grease passage way blocked by the protective plastic membrane not being removed prior to assembly.
The DBBC has had new filters installed to the input IF's for the L-band; this will allow the RF to be received directly with no mixing. 
We are working on finishing the L-band backend for LBA sessions in 2013.
We have undertaken a big tidy up of the racks to try and make the systems more manageable; all systems not required in the 12m system
racks have been moved to the 30m computer room across the road to help reduce the heat sources within the 12m control room, as the maser resides here.

\subsection{30m: Progress and Issues}

\begin{figure}[!h]
\centering
\begin{tabular}{cc}
\includegraphics[height=5cm, bb=0 0 1536 2048]{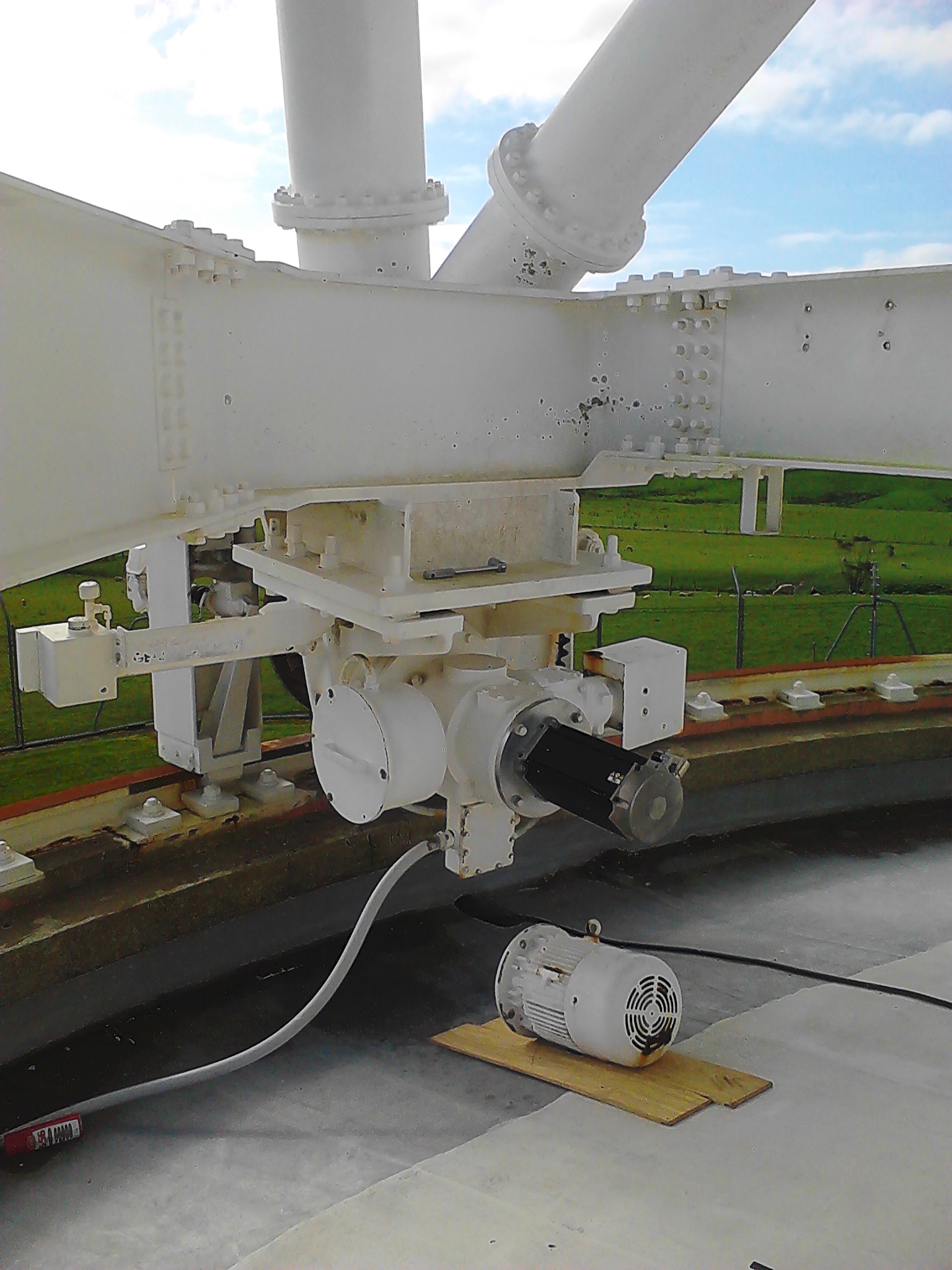}&\includegraphics[width=6cm, bb=0 0 2048 1536]{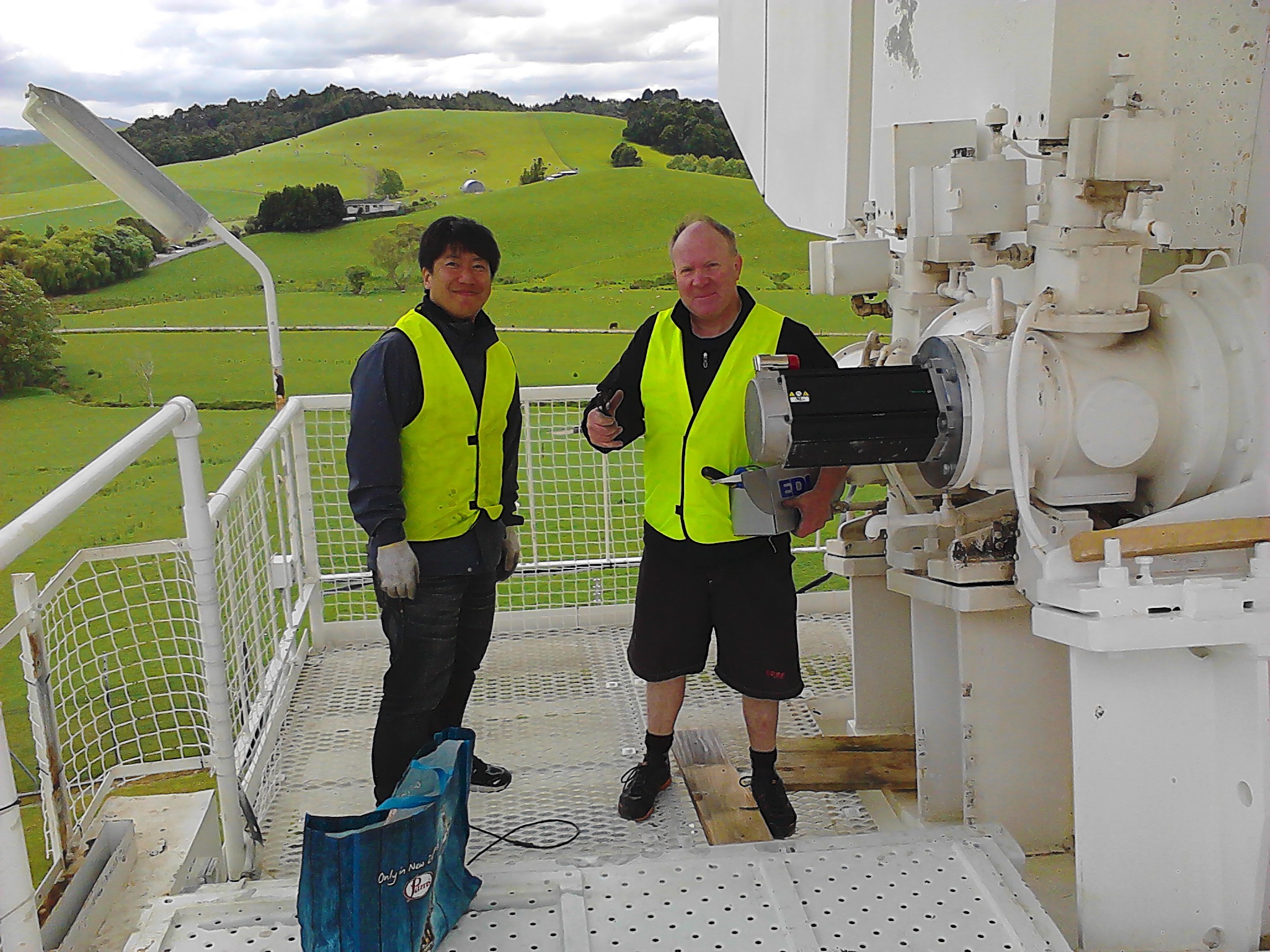}\\
\includegraphics[width=6cm, bb=0 0 3648 2736]{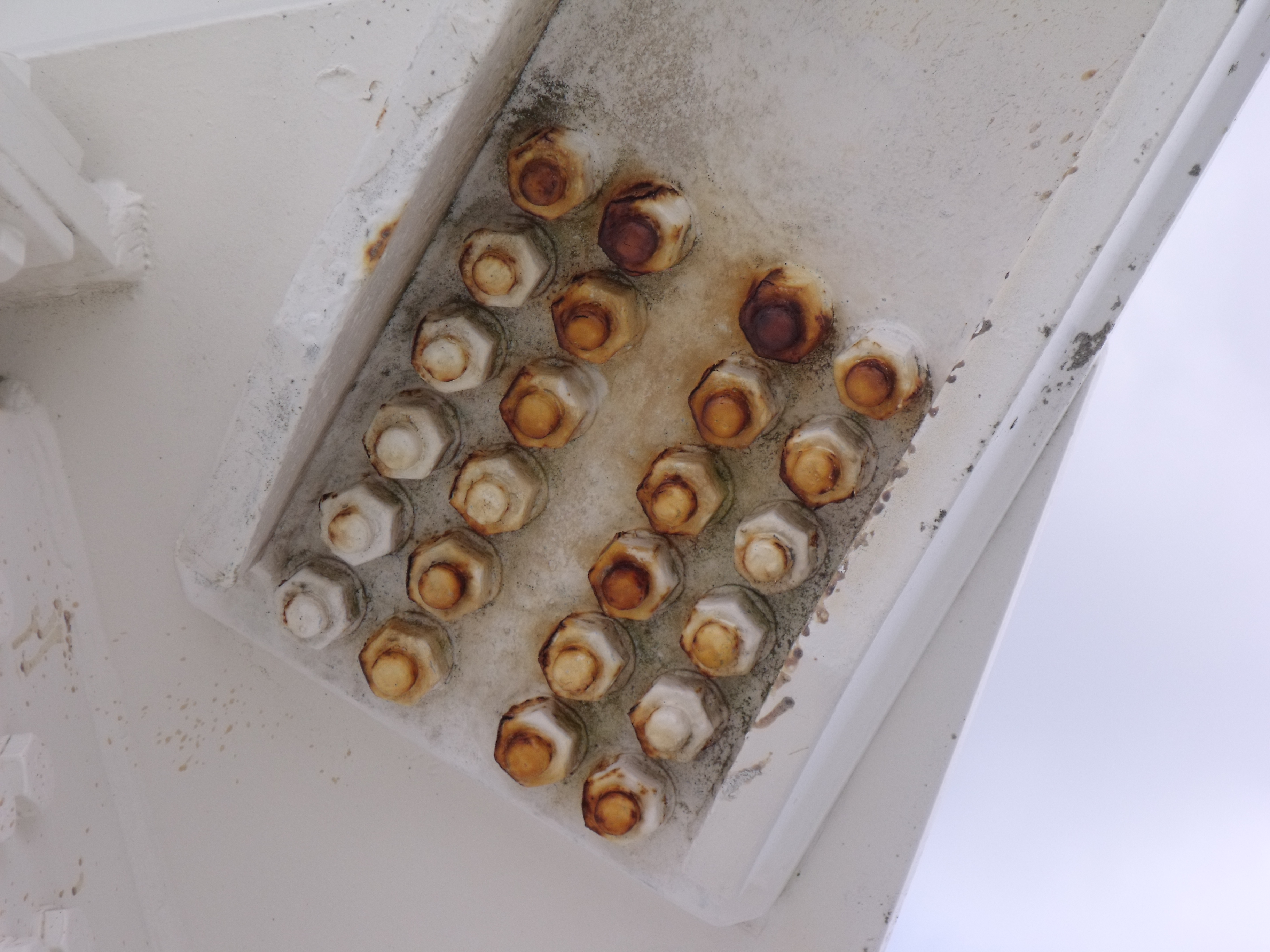}&\includegraphics[height=5cm, bb=0 0 1536 2048]{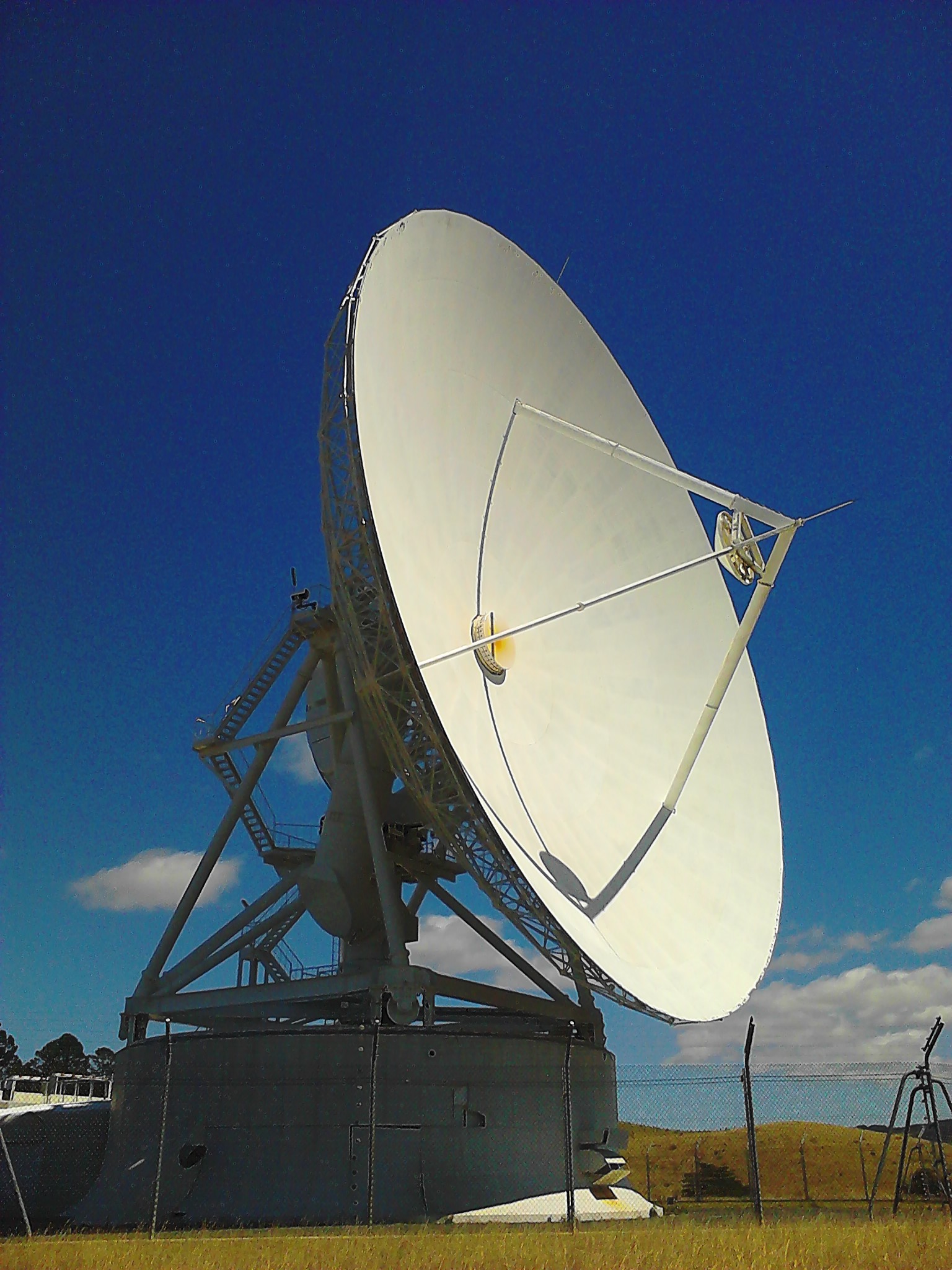}\\
\end{tabular}
\caption{The 30m, Top-left one of the old azimuth motors being replaced. Top-right Tim and Takiguchi having finished replacing one of the elevation motors. Bottom-left an example of the corrosion we have to get on top off. Bottom-right with the new motors and control system we are now able to move the dish once again. Image Credit: S. Weston}
\label{fig:30m}
\end{figure}

The 30-m Cassegrain beam-waveguide NEC antenna was used by Telecom NZ after construction in 1984. It was handed over to AUT in 2010. 
Renovation of the 30-m started in 2011. Old Azimuth and elevation motors have been removed and replaced with new ones (see Figure \ref{fig:30m}). A new control system from Control Techniques, UK, very similar to the system used on the 12m was installed. A new cable wrap system has been installed and new cables pulled through. In mid February 2013 Mark Godwin will be on-site to commission the new control system.
A bolt replacement program by a local rigging contractor has started. We are in the process of negotiating a maintenance contract with them for both the 30m and the 12m.
Additional DBBC and Mark 5C have been ordered for the 30m.
Work continues on using the existing satellite receiver in C-Band, but we are also talking with other parties about receiver systems. Obviously we would like to go L,S,C and X but are unsure of the waveguide optics.

\subsection{Warkworth Network}

We continue to e-transfer data to the correlators. We have installed a new 10Gbit switch in the 30m computer room. This provides fibre point to point connectivity between the DBBC and Mark 5C via the Fila10G interface and gives us a 10Gbit backbone at the observatory. The network topography was changed so that the 12m is now a spur from the 30m KAREN POP. Warkworth now has a direct IP presence on KAREN, so data transfers should no longer go through the University Campus any more (of mutual benefit to us and campus LAN users). It is hoped with the upgrades taking place with the KAREN network that we will have 10Gbps international connectivity to/from the observatory in the near future.
The 30m building has been wired cat6 and fibre has been installed from the computer room through to the pedestal room beneath the dish where the wave guide terminates.
In addition equipment has been installed for distribution of 1PPS and 10MHz from the maser at the 12m over to the 30m pedestal room using a Symmetricom RF via fibre system.

%\section{Staff}
%%Section 3: Staff. Please provide a list of your staff members who are contributing to IVS, provide some information about %them and indicate what each person is working on.

%\vspace{-2cm}
%\begin{wraptable}{r}{95mm}
% \begin{center}
%%\begin{table}[!h]
%%   \centering
%%   \par\medskip\par
%   \begin{tabular}{|l|l|}
%   \hline
%        Sergei Gulyaev & Director \\
%        Tim Natusch & Deputy Director \\
%        Stuart Weston & software, technician, observer  \\
%        Hiroshi Takagushi & principle investigator \\
%        Lewis Woodburn & site maintenance \\
%        Patricia Sallis & administration \\
%   \hline
%   \end{tabular}
%%\end{table}
%\vspace{-1cm}
% \end{center}
%\end{wraptable}

%\vspace{1.5cm}
%The personnel associated with the geodetic VLBI program at Warkworth and their primary responsibilities are:

%\vspace{0.3cm}
\section{Current Status and Activities}
%Section 4: Current Status and Activities. Please provide information about the current status and activities of your %component during the reporting period, such as operational support, milestones reached, analysis performed or ongoing %projects. Please describe any significant changes compared to the previous year's report, such as staffing changes, new %equipment, changes in methods, new types of results or a change in your activity level. You do not need to repeat %information already provided in the previous year's report.
In March 2012 we had a visit by Gino Tuccari with Jim Lovell and Jamie McCallum from Hobart for a mini workshop on the DBBC looking at diagnoses and issue resolution. Later in the year during September Ed Himwich also paid a visit, during which he installed the new Field System DBBC support components. Now when we drug the generated schedule prc file contains the appropriate commands to program and setup the DBBC.
We have successfully installed SDK 9.2 on our Mark 5B to support diskpacks larger than 8TB. We now have a station pool of five 16TB diskpacks which we will be using for LBA and the AUSTRAL experiments with AuScope antennas in Australia. 
%We continue to e-transfer data to the correlators.
In addition to the IVS and LBA observations the WARK12M is now also a tracking station for SpaceX on their supply missions to the International Space Station.

\begin{figure}[htb!]
\begin{center}                   % center environment
  %\parbox{20cm}
  {
%  \epsfclipon
%  \epsfxsize=16cm
%  \epsffile
  %[14 14 783 591]
%  {nswark0x.pdf}
  \includegraphics[scale=0.7]{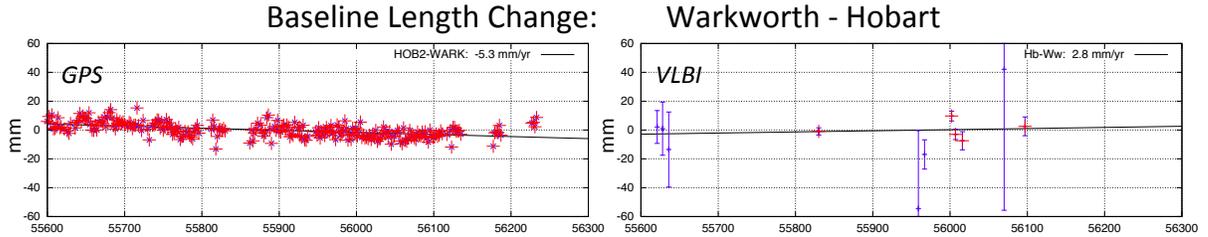}
  }
\caption{The baseline length changes of Warkworth-Hobart baseline derived from both GNSS and VLBI data sources for the epoch 2011-2013. GNSS results were produced by analysis of the GNSS data (IGS: alic, auck, hob2, karr, kat1, yarr + PositioNZ: wark) using the GAMIT/GLOBK software package. VLBI baseline behaviour was determined from results produced by the IVS analysis center.  The difference of colour indicates whether these data were used (red) for the calculation of the change rate or excluded (blue).}
\label{fig1}
\vspace{-0.8cm}
\end{center}
\end{figure}

\subsection{Co-operative observation}

Co-operative observation for geodetic purposes with NICT (Japan) and University of Tasmania (Australia) started. The baseline of WARK12M and KASHIM11 is a long north-south baseline of over 8,000-km. The first observation of the WARK12M-KASHIM11 baseline was carried out in April 2012 and the baseline length has been determined as 8,075,003,545$\pm$150 mm. By repeating the observation, we expect to obtain information about the relative tectonic motion of Japan and New Zealand. Also, we are working to establish the ability to derive EOP ultra-rapidly by utilising this baseline, existing UT1 products (such as from the IVS INT2 sessions) and data transfers by high-speed network. The University of Tasmania operates three 12-m radio telescopes located in Hobart (HOBART12), Yarragadee (YARRA12M), and Katherine (KATH12M) under the AuScope project \cite{lovell2013}. These three telescopes and WARK12M are located on the Australian tectonic plate and consequently are ideally placed for measurements of intra-plate deformation. WARK12M-AuScope observations started in July 2012. Monthly 24-h observations and a series of multi-day observations together with AuScope antennas will be scheduled from early 2013.

Figure \ref{fig1} shows the baseline length changes of Warkworth-Hobart baseline derived from both GNSS and VLBI data sources for the epoch 2011-2013. We checked all of the baseline of AuScope and WARK12M. GNSS results indicate a small intra-plate deformation. On some baselines current VLBI results appear to be in disagreement with the GNSS data. At this stage it is not possible to comment further on this discrepancy due to the statistically small number of VLBI observations. Our proposed co-operative observations will add vital data points, reduce the noise and compensate for this current VLBI weakness.

%\section {Future Plans}
%%Section 5: Future Plans. Please describe your plans for the coming year: scheduled operations, planned activities, %maintenance, upgrades, new equipment, new staff, new software development or algorithms, new hardware arriving, etc.

%This year we will configure and test the DBBC Fila10G to Mk5C over our optical network via 10Gb switch, this will provide more flexibility in recording.
%The AUSTRAL experiments started \cite{lovell2013} in 2012 will be expanded, also we hope to have a more active participation in the IVS observing schedule becoming a more reliable station.
%We also continue to promote and try to educate the New Zealand Geodetic community to the benefits of obtaining a second 12m dish in South Island, NZ which would lie on the south side of NZ fault and not on the Australian plate.
%Work will continue on the commissioning of the 30m dish.

\end{document}